\RequirePackage{ifpdf}
\documentclass[12pt]{JHEP3}
\pdfoutput = 1

\usepackage{color}

\usepackage{multicol}

\usepackage{epsfig,bm}
\epsfclipon
\usepackage{amsmath,amssymb,amsbsy,amstext,amsthm}

\let\normalcolor\relax  

\newcommand{\roughly}[1]{\mathrel{\raise.3ex\hbox{$#1$\kern-0.85em
\lower1ex\hbox{$\sim$}}}}

\newcommand{\lsim}{\roughly<}
\newcommand{\gsim}{\roughly>}

\def\be{\begin{equation}}
\def\ee{\end{equation}}
\def\ba{\begin{eqnarray}}
\def\ea{\end{eqnarray}}

\usepackage{ulem}

\theoremstyle{definition}

\title{Can non-local or higher derivative theories provide alternatives to inflation?}

\author{Ghazal Geshnizjani${}^{1,3,4}$, Nahid Ahmadi${}^{2, 3}$\\
$^1$Department of Applied Mathematics, University of Waterloo, Waterloo, Ontario, N2L 3G1, Canada\\ $^2$Department of Physics, University of Tehran, Kargar Avenue North, Tehran 14395-547, Iran\\$^3$ Perimeter Institute for Theoretical Physics, 31 Caroline St. N., Waterloo, ON, N2L 2Y5, Canada\\$^4$Department of Physics, University of Waterloo, Waterloo, Ontario, N2L 3G1, Canada}

\date{}

\abstract{The standard mechanism for producing the observed scale-invariant power spectrum from adiabatic vacuum fluctuations relies on first order derivative of fields in the action for curvature perturbations. It has been proven \cite{Geshnizjani:2011dk} that, under this ansatz, any theory of early universe that matches cosmological observations should include a phase of accelerated expansion (i.e. inflation) or it has to break at least one of the following tenets of classical general relativity: Null Energy Conditions (NEC), subluminal signal propagation, or sub-Planckian energy densities. We extend this proof to a large class of theories with higher (spatial)  derivative or non-local terms in the action. Interestingly, only theories in the neighborhood of Lifshitz points with $\omega_k \propto k^0$ and $k^3$ remain viable.  }

\preprint{}

\begin{document}
\section{Introduction}
Observations of Large Scale Structure (LSS) of the universe \cite{Reid:2009xm} and Cosmic Microwave Background (CMB) \cite{Ade:2013zuv} indicate an approximate gaussian and  scale-invariant  power spectrum of primordial perturbations.  These features persist over at least three decades of wavelengths, from largest to smallest observable scales in sky. Tracking the evolution of the universe back in time,  these modes would correspond to scales larger than Hubble radius, at early times.  Inflation, which is an early phase of quasi-deSitter background expansion, provides a compelling mechanism to generate a gaussian and scale-invariant power spectrum from vacuum quantum fluctuations, while at the same time, stretching them to super-Hubble scales. In spite of its success in fitting observantional data, testing inflation with certainty remains a difficult task. This is because without (a constrained) realization of Inflationary action in any fundamental theory, phenomenologically, different inflationary scenarios can accommodate a variety of possible predictions \cite{Bean:2008ga}. While we cannot prove or rule out inflation with certainty, we can aim to confirm it by exclusion principle, i.e. addressing whether there can be another compelling alternative theory of early universe that is consistent with data. In fact, there have been many proposals over the past two decades, exploring other alternatives. These proposals range from generating perturbations during ordinary expanding backgrounds with rapidly varying  speed of sound \cite{ArmendarizPicon:2003ht, Magueijo:2008pm, Bessada:2009ns, piao2007seeding}, to bouncing (see \cite{Khoury:2011ii} and references there) and static universes \cite{Creminelli:2010ba, liu2011galileon}, varying fundamental speed of light \cite{Albrecht:1998ir} or to theories without explicit space-time backgrounds \cite{Nayeri:2005ck}. Interestingly, for any of these proposals to succeed, at least one of the main tenets of general relativity have to be broken. The list usually includes violating {\it Null Energy Condition} (NEC), going through a singularity or superluminal speed of sound or light. A theorem was recently proven in \cite{Geshnizjani:2011dk} that shows why this is true for any alternative scenario to inflation. The theorem states that any scenario that does not have a phase of accelerated expansion must satisfy one or more of the following conditions: violating Null Energy Conditions (NEC),  superluminal speed of sound, or super-Planckian energy densities. 
  This result was based on assuming standard mechanism for producing the observed scale-invariant power spectrum from {\it adiabatic} vacuum fluctuations, where the effective mass mimics the time dependence of effective mass in de Sitter background. What is interesting about this result is that, it is {\it not} the intrinsic nature of generating super-Hubble fluctuations, but rather the fact that  they persist over at least three decades that makes accelerated expansion inevitable.
  
There are other studies investigating specific models \cite{Linde:2009mc, Easson:2013bda} or examining the strong coupling regimes which reach almost similar conclusions \cite{Joyce:2011kh, Baumann:2011dt}.  While these methods are complementary  to each other, they differ from each other in some aspects. For example, the advantage of the above theorem is that it is more definitive in constraining scenarios that do not involve early time acceleration. 
However, the advantage of the strong coupling argument is that it can be extended to bouncing models which by nature have to go beyond the validity of General Relativity to cross the bounce.  
 
Since this proof relies on considering a standard action for curvature perturbations with only first order derivatives of fields,  as a natural next step one can ask  whether this theorem still holds if higher derivative or non-local terms contribute to generation of power spectrum\footnote{For other applications of non-local actions in cosmology, such as explaining the late time acceleration of universe or resolving big bang singularity see \cite{Deser:2013uya, biswas2010towards}.}. Here, we investigate this very question, i.e. whether this proof is robust against including non-trivial terms in the action such as those appearing in ghost inflation \cite{ArkaniHamed:2003uz}. While we are carrying a comprehensive analysis of a large range of possibilities, we are leaving out some unusual ones. These are cases that are very unlikely to lead to a near scale invariance power-spectrum, but are also hard to rule out definitively without knowing more details about the actual dynamics in the scenario.  For instance, one may envisage three or more terms in the action controlling the dynamics of perturbations in different regimes, conspiring to produce a scale-invariant power spectrum. Ignoring such possibilities, we restrict to actions with only two dominant terms (at quadratic level in amplitude of perturbations) and will keep the arguments independent of specific details of the background evolution. 

Furthermore, we only include terms that are higher order in spatial derivatives and not time.  Actions with higher time derivatives either  suffer from Ostrogradski instabilities or are secretly not higher order (degenerate actions)  \cite{Ostrogradski, deUrries:1998bi, Woodard:2006nt}. 

The format of the paper is as follows: in Section (\ref{setup}) we review the proof derived in \cite{Geshnizjani:2011dk} with additional  improvements on the argument. In Section (\ref{higherspatiald}) we investigate extending this proof by including pure spatial higher order derivative or non-local terms in the action. In (\ref{time spatial}) we discuss including terms that involve both time and higher order/fractional spatial derivatives which lead to some peculiar behaviours. We make our concluding remarks in Section (\ref{conclusion}).  We also recommend the Appendixes (\ref{generalbg}) and (\ref{apxHorizonz}) for readers who find derivation of scale-invariance in non-inflationary backgrounds or behaviour of different cosmological horizons in such backgrounds counter intuitive.


\section{Necessary conditions to produce the power spectrum from a standard quadratic action} \label{setup}
To calculate the power spectrum of inhomogeneities in our universe,  it is sufficient to perturb the action around a flat Friedmann-Roberston-Walker (FRW) background, up to quadratic order in curvature perturbations. We start by writing the most general action for curvature perturbations, including only the quadratic terms and restricting to first order derivatives:

\be \label{action1}
S_2={M_{pl}^2\over 2}\int d^3x~ d\tau~ z(\tau)^2 \left [ \left ({\partial\zeta\over \partial\tau}\right)^2- c_s(\tau)^2(\nabla\zeta)^2 \right ]. 
\ee
Here $\tau$ is the conformal time, $c_s$ can be interpreted as speed of sound and $z(\tau)$ is some time dependent function that can be derived from a full action and depends on the homogenous solution for background. As long as a cosmological model has only one degree of freedom, the above action captures the most generic features of the cosmological perturbations.  One can show that for hydrodynamical fluids and scalar fields $z(\tau)$ satisfies \cite{Brandenberger:1993zc, Garriga:1999vw} : 
\be
 z\equiv {a\sqrt{2 \epsilon}\over c_s}.  \label{z}
 \ee
For our discussion here, the actual dependence of $z$ on background parameters is not relevant. So we just assume it is a differentiable  function of time. 

The reason for not including a mass term such as $m^2(\tau)\zeta^2$ (or a potential term $V(\zeta)$\footnote{Any constant term that does not depend on $\zeta$, is assumed to have been included in the zeroth order terms of the action.}), in the action is that, we are requiring $\zeta\rightarrow constant$ to be a solution in the infrared limit. This is a necessary condition in order to  obtain an attractor solution\footnote{The sufficient condition is that the constant mode of $\zeta$  be the dominant mode as well.}.   

 Even though the overall sign of the action does not change our argument, we expect the $z^2>0$ in most scenarios. Choosing a negative sign could potentially lead to ghost instabilities in the presence of other fields. 

The only remaining possible corrections to second order terms that can contribute to power spectrum are terms including higher  derivatives or non-trivial operators. Here we review the standard mechanism for generating scale-invariant curvature perturbations based on the Action (\ref{action1}) and will consider cases where such terms are important in the following sections. 

We now follow the framework introduced in  \cite{Khoury:2008wj} to transform the action (\ref{action1}) into an action with unit speed of sound. This can be accomplished using the time transformation $dy=c_s d\tau$:
\be \label{action2}
S_2={M_{pl}^2\over 2}\int d^3x~ dy~ q^2 \left [ \left ({\partial\zeta\over \partial y}\right)^2- (\nabla\zeta)^2 \right ], 
\ee
where  
\be\label{q}
q\equiv z \sqrt{c_s}.
\ee 
The next step is applying the standard mechanism for quantization of a scalar field in curved space-time. This task is often conducted by transforming the action into a canonically normalized action for a new auxiliary field $v\equiv M_{pl}~ q ~\zeta$ in a flat background and with time dependent mass. The Action (\ref{action2}) is then rewritten as
\be \label{action3}
S_2={1\over 2}\int d^3x~ dy \left [ v^{\prime2}- (\nabla v)^2 +{q''\over q}v^2\right ], 
\ee
where prime represents $\partial/\partial y$. The field $v$ can now be quantized by being promoted into the field operator $\hat{v}$ and imposing commutation relation on $\hat{v}$ and its canonically conjugate momentum. 

Next, we proceed by setting the  initial conditions to  adiabatic vacuum.
The equation of motion for $v$ in Fourier space is given by 
\be \label{eqnforvk}
v_k''+\left(k^2-{q''\over q}\right)v_k=0. 
\ee
Therefore, the leading WKB condition reduces to 
\be
\omega_k^2 \gg \left|{\omega_k'' \over 2\omega_k}-{3\omega_k'^2\over 4\omega_k^2}\right|,
\ee 
where $\omega_k^2 \equiv k^2-q''/q~$\footnote{$\omega_k$ can be real or imaginary.}. In the regime that the WKB approximation holds, the WKB solution
\be \label{vacuum}
v_k={1\over \sqrt{ \omega_k}}\exp\left[-i \int^y \omega_k(y)dy\right]
\ee
corresponds to the adiabatic vacuum \cite{Birrell:1982}. 

The standard mechanism for generation of scale invariant spectrum assumes that the time dependence of $q''/q$ is such that early times coincide with $k^2\gg |q''/q|$, and also that the WKB approximation holds.  In this limit, the solutions asymptote to simple oscillatory behaviour. At late times, WKB approximation often breaks down in order to result in a scale-inavriant spectrum (which can be interpreted as particle production, if it precedes a second adiabatic regime).  Mathematically, this implies $|q''/q|$ is a growing function in $y$ and also it is a smooth function such that its time derivatives do not diverge at early times. In other words, as $(y-y_{end}) \rightarrow - \infty$, we recover  $|q''/q|\ll k^2$, $|q''/q|^\prime \ll k^3$, etc. For example, for any scenario in which $|q''/q|$ has very sharp spikes, then we could have regimes where  $|q''/q|\ll k^2$ but $|q''/q|^\prime \gg k^3$ and the WKB approximation is not valid.

\begin{figure}[htpb]
\includegraphics[width=\columnwidth]{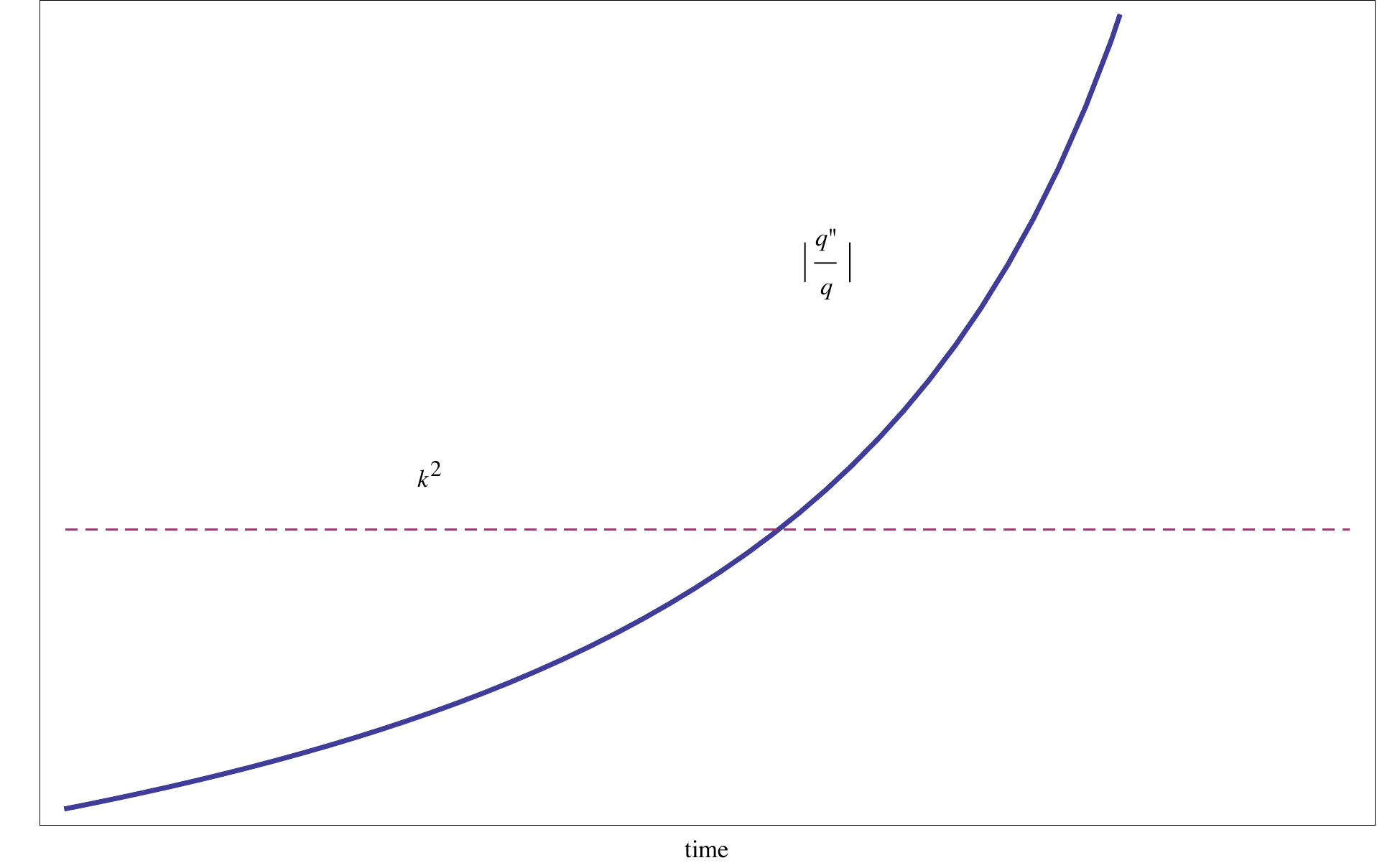}
\caption{A smooth function for $|q''/q|$ that vanishes in early times.}\label{mt}.
\end{figure}

Therefore, we restrict our models to cases where  $|q''/q|$ increases in time smoothly. This enables us to impose adiabatic vacuum condition such that  every mode starts in an approximately Minkowski vacuum state at early times. At late times, modes evolve such that $k^2 \ll q''/q $ and the general solution to Equation (\ref{eqnforvk}) should asymptote to the following form: 
\be
v_k\sim c_1 q(y)+ c_2~ q(y)\int^y \frac{1}{q^2(\tilde{y})} d\tilde{y}
\ee
It is clear here that the asymptotic solution will lead to conservation of $\zeta$ or an attractor solution for a FRW universe only if the time dependence of $\int d\tilde{y}/{q^2(\tilde{y})} $ decays away in time. Therefore $|q(y)|$ has to be either growing in time or decaying slow enough to avoid even a logarithmic divergence. 
Satisfying this condition in addition to $|q''/q|$  being a uniformly growing function of time restricts the choices of allowed functions for $q(y)$ significantly. 

A simple choice that automatically satisfies both of these conditions is to assume $ q\propto (y_{end}-y)^\mu$ with $\mu<1/2$.  We can choose $y_{end}=0$ for convenience and since $dy > 0$, the  range of the variation for $y$ will be $y \in [-\infty , 0]$.  Now substituting this assumption in to Eq. (\ref{eqnforvk}), we arrive at: 

\be \label{vkqmu}
v_k''+\left[k^2-{\mu(\mu-1)\over y^2}\right]v_k=0. 
\ee

The solutions to this equation are of the form: $C~\sqrt{|y|}~f_\nu(ky)$ where $f_\nu$ is a Bessel function of order $\nu$
\ba \label{defnu1}
\nu&=&
|\mu-1/2|.
\ea
  Therefore, we can write the general solution as: 
\be
v_k(y)=C_1 \sqrt{y}H^{(1)}_\nu(ky)+C_2 \sqrt{y}H^{(2)}_\nu(ky),
\ee
where $H^{(1)}_\nu$ and $H^{(2)}_\nu$ are Hankel functions of the first kind and second kind respectively. 
The asymptotic behaviour of Hankel functions for $|ky|\gg |\nu^2-1/4|$ (after discarding a constant phase) are: 
\ba
H^{(1)}_\nu(ky)&\sim& \sqrt{2\over \pi k|y|}~\exp{[-i ky]}\\
H^{(2)}_\nu(ky)&\sim &\sqrt{2\over \pi k|y|}~\exp{[i ky]}
\ea
where we also used the fact that $y<0$. However, at early times when $k^2\gg |q''/q|$ (or $ky \gg |\nu^2-1/4|$) we are also  imposing the adiabatic vacuum condition from Eq. (\ref{vacuum}): 
\be
v_k(y)\sim \sqrt{{1\over k}}\exp{[-i ky]}.
\ee
This implies for our general solution that $C_2=0$ and $C_1=\sqrt{\pi\over 2}$. We can now calculate the spectrum by looking at the asymptotic behaviour of $H^{(1)}_\nu(ky)$ in the limit that  $|ky|\ll 1$: 
\ba
 {\cal P}^\zeta_k&\equiv& k^3{ |v_k^2|\over M_{Pl}^2 q^2} \nonumber \\
&=&k^3 \left\{ {\pi \over 2} y [H^{(1)}_\nu(ky)]^2\right\}\left({ 1\over M_{Pl}\alpha y^\mu}\right)^{2}\nonumber\\
&\simeq& k^3 \left[ {\pi \over 2 M_{Pl}^2 \alpha^2} \right]y^{1-2\mu}\left[{\Gamma(\nu)\over \pi}\left({2\over k y}\right)^\nu\right]^2, ~~~{\rm for} ~~~|ky|\ll 1 \label{PS1}
\ea
where $\alpha$ is the constant of proportionality for $q\propto |y|^\mu$. 
The above result yields that 
\be 
n_s-1=3-2\nu.
\ee
Thus, to have a near scale invariant spectrum $\nu$ has to satisfy: 
\be
\nu\simeq {3\over 2}. 
\ee
We can now use Eq. (\ref{defnu1}) and obtain two solutions for $\mu$:
\ba
\mu_1&=&{1\over 2}+\nu={1\over 2}+{3\over 2}=2,\\
\mu_2&=&{1\over 2}-\nu={1\over 2}-{3\over 2}=-1.
\ea
However, as we pointed out before, only a value of $\mu< 1/2 $ will lead to conservation of $\zeta$, i.e. an attractor solution.  Therefore, any cosmological scenario with approximate $q\propto 1/y$ can lead to a scale invariant spectrum. This result not only alleviates the need for a slow-roll inflation in order to generate scale-invariant  fluctuations, but in fact proves that one can obtain a scale invariant spectrum in any expansion history (see Appendix \ref{generalbg}).  In other words , even though applying the naive slow-roll result $n_s-1\sim 2(\epsilon+\frac{d \ln{\epsilon}}{Hdt}+\frac{d \ln{c_s}}{Hdt} )$ may suggest  otherwise, calculating  the spectral index $n_s$ would result in 
\be
n_s-1=0.
\ee
In practice, to obtain a non-zero tilt compatible with observations, $q$ must have deviations from the above solution. For example, allowing for $\mu$ to be different from $-1$ or have small time dependence can lead to a small tilt \footnote{From theoretical perspective there is no obstacle in constructing cosmological models with exactly scale-invariant spectrum \cite{Starobinsky:2005ab}.}.  For more discussion on 
this and the generalized slow-roll condition see \cite{Hu:2011vr}.

 It is crucial to recognize that, in general, the scale which sets the mode freezing is not the same as when the mode crosses the Hubble radius. We will use the following notations to differentiate between these two scales: 
 \be
R_H\equiv \frac{1}{a H}, 
\ee  \label{Rh}
\be  \label{Rz}
R_\zeta \equiv \sqrt{ {q\over q^{\prime\prime}}}\sim - y, 
\ee
where, $R_H$ is referred to as comoving {\it Hubble} radius/horizon while  $R_\zeta$  denotes comoving ``freezout"  radius/horizon. 

We refer the readers, to Appendix \ref{apxHorizonz} for a review of different cosmological scales. Our derivation so far postulates that freezout radius and not Hubble radius should be shrinking to generate perturbations. In what follows we demonstrate why the shrinkage of comoving Hubble radius, i.e. acceleration of expansion is still a relevant mechanism to explain the observations. 

We know that during the matter and radiation eras, Hubble radius has been growing so the large scale perturbations we observe today were outside of the  {\it Hubble} horizon in the past.  The wavelength of these perturbations range from $\lambda_f\sim 1$ Mpc to $\lambda_i\gsim 3 \times 10^3$ Mpc. A simple estimate yields that  $R_H\sim 1$ Mpc at redshifts around $z_*\sim 4\times 10^5$ or temperatures around $T_*\sim 0.1$ keV and perturbations with wavelengths\footnote{scales corresponding to near scale invariant power spectrum.}  $\lambda \gsim 1$ Mpc, were all larger than $R_H$ at that time. This observation, together with the observational constraints from Big Bang Nucleosynthesis  (BBN), which require the universe to have become radiation dominated by  $T\sim1$ MeV ($z\sim 4\times 10^{9	}$) impose significant constraints on applying above mechanism for generating perturbations \cite{Geshnizjani:2011dk}. 
The consequence is that  any scenario that does not have a phase of accelerated expansion (or inflation) must satisfy one or more of the following conditions: violating Null Energy Conditions (NEC),  superluminal speed of sound, or super-Planckian energy densities. 

A brief summary of the derivation is provided here, as the extension of this argument to more general actions bears much resemblance to the standard case. 
The proof starts by assuming that the expansion of universe has neither violated NEC nor has it gone through an accelerated phase in the past. This assumption can be expressed as $\epsilon \geq 1$ in early universe where $\epsilon$ is defined as $\epsilon\equiv-\dot{H}/H^2$ and it automatically implies that $R_H$ has always been growing in time.  

 Since the generation cannot happen during the radiation era, every mode must have exited the freezeout horizon before BBN, and according to Eq. (\ref{Rz}) at a time corresponding to $y \sim 1/k \sim \lambda$, in order to become scale invariant.  The wavelength $\lambda_i$ corresponding to the largest scale we observe in the sky, is also at least  three orders of magnitude larger than $\lambda_f$.  Summing up all these arguments,  the following conditions must be satisfied: 
\ba \label{ineq0} 
|y_i|&\gsim& 3\times 10^3 |y_f| \\
|y_f|&\sim& \lambda_f\sim R_H(z_*)\sim10^4 R_H(z_{BBN}), 
\ea
which leads to 
\be \label{ineq1}
{y_f-y_i\over R_H(z_{BBN})}\gsim 3\times 10^7.
\ee
Next, integrating the continuity equation,
\be
{\dot{\rho}\over \rho}=-2\epsilon H, 
\ee
over time, $t$, we arrive at 
\ba \label{densitytime}
\ln{{\rho_i\over\rho_f}}&=&2\int_{t_i}^{t_f} \epsilon H dt \nonumber\\
 &>& 2 R^{-1}_H(\tau_f)\epsilon_{\rm min} (\tau_f-\tau_i).
\ea
Since  $R_H(y)<R_H(y_f)\lsim R_H(z_{BBN})$ for any $y<y_f$, we obtain
\be \label{ineq2}
\ln{\rho_i\over \rho_f}>\left ({2\epsilon_{min} \over R_{H}(z_{BBN})}\right) {y_f-y_i\over \bar{c}_s}, 
\ee
where $\bar{c}_s$ has been defined as $\bar{c}_s\equiv (\int^{\tau_f}_{\tau_i}c_sd\tau)/(\tau_f-\tau_i)= (y_f-y_i)/(\tau_f-\tau_i)$. 
 Combining inequalities (\ref{ineq1}),  (\ref {ineq2}) and $\epsilon \geq 1$ leads to
\be \label{ineq3}
\ln{\rho_i\over \rho_f}> {6 \times 10^7 \over \bar{c}_s}.   
\ee
We can rearrange this condition as 
 \be \label{ineq4}
\bar{c}_s\left[ 88\ln10 + \ln{\rho_i\over M^4_{pl}}-\ln{\rho_f\over \rho_{BBN} }\right]> 6 \times 10^7. 
\ee
 Since $\rho_f\gsim \rho_{BBN}\sim (1~{\rm MeV})^4$, for above inequality to hold, one of the following cases must be true: 
 \be
 \rho_i> 10^{26057581} M^4_{pl} ~~~~~~~({\rm if}  ~\bar{c}_s \leq 1),
 \ee
 or 
 \be
 \bar{c}_s>  3\times 10^5~~~~~~~({\rm if}   ~\rho_i \leq M^4_{pl} ).
 \ee
Both of these inequalities, i.e. {\it super-Planckian} energy density
 or {\it superluminal} propagation speed,  point to a breakdown of general relativistic physics\footnote{The argument presented here is an improved version of the proof presented in \cite{Geshnizjani:2011dk}}. 
 


\section{Including non-standard spatial derivatives in the action}\label{higherspatiald}
We now proceed to investigate whether the proof presented in the last section can be extended to actions where non-standard spatial derivative terms have a dominant role.  We start by replacing the term $(\nabla\zeta)^2$ in action (\ref{action1}), with terms of type $(\nabla^m\zeta)^2$ with $m$ being an non-zero integer number:

\be \label{actionhighergrad1}
S_2={M_{pl}^2\over 2}\int d^3x~ d\tau~ z(\tau)^2 \left [ \left ({\partial\zeta\over \partial\tau}\right)^2- \bar{\lambda}^{2m-2}b(\tau)^2(\nabla^m\zeta)^2 \right ]. 
\ee
Here,  $b(\tau)^2$ is  a possible time dependent coefficient and we have introduced the constant scale,  $\bar{\lambda}$ to keep $b(\tau)$  dimensionless. The assumption of a positive $b(\tau)^2$, as opposed to a negative function, has been presumed to allow for a well behaved WKB limit in early times, and is similar to discarding the standard quadratic action with imaginary sound speeds. 
We can re-express this action in Fourier space\footnote{Our crucial assumption here is that the perturbations have well defined Fourier transformations.} and then  extend the scope of our analysis to include non-integer values of $m$ as well,
\be \label{actionhighergrad2}
S_2={M_{pl}^2\over 2}\int d^3k ~d\tau~ z(\tau)^2 \left [ \left ({d\zeta_k\over d\tau}\right)^2- \bar{\lambda}^{2m-2}b(\tau)^2k^{2m}\zeta_k^2 \right ]. 
\ee
 In general, non-positive and non-integer values of $m$ can be associated with integration operators or fractional derivatives\footnote{Fractional derivatives and integrations can be defined as,
\ba
\nabla^m\zeta&\equiv&\frac{1}{(2\pi)^3}\int_{-\infty}^{+\infty}\left(-ik\right)^{m}\zeta_k e^{-ik.x}d^3k \nonumber\\
&=&\int_{-\infty}^{+\infty}D^{m}\left(x-y\right)\zeta\left(y\right)dy,\nonumber
\ea 
where 
\be
D^m(x-y)\equiv\frac{1}{(2\pi)^3}\int_{-\infty}^{+\infty}(-ik)^{m}e^{-ik.(x-y)}d^3k.
\ee
In general $D^m(x-y)$ does not vanish for nonzero values of $(x-y)$ which yields to nonlocality. However, when $m$ is a positive integer number, $D^m(x-y)$ corresponds to $m$th derivative of Delta function, $\delta^3(x-y)$, and restores locality.}. These operators, unlike regular derivatives, will no longer be local operators. Surprisingly, as we see below, our analysis, in spite of its simplicity, is able to rule out most of this parameter space as well. 
We could also assume a more general class of operators by replacing $\bar{\lambda}^{2m-2}b(\tau)^2k^{2m}$ with an arbitrary function $f(k)$. However, as we will see only those operators corresponding to powers of $k$ could generate a scale invariant power-spectrum.

Once again we can reparametrize time to absorb the time dependence of $b(\tau)$ into the variable $y$ defined as $dy\equiv b(\tau)d\tau$ and function $q(y)\equiv z\sqrt{b}$.  This will turn the action into the following form,

\be \label{actionhighergrad2}
S_2={M_{pl}^2\over 2}\int d^3k~ dy~ q(y)^2 \left [ \left ({d\zeta_k\over dy}\right)^2-\bar{\lambda}^{2m-2} k^{2m}\zeta_k^2 \right ], 
\ee
where $y$ has the dimension of time. We can now proceed to introduce the canonical variable  $v_k\equiv M_{pl} ~q ~\zeta_k$ similar to previous section and express the action in a canonical form
\be \label{actionv}
S_2={1 \over 2}\int d^3k ~dy \left [ v_k^{\prime2}-\bar{\lambda}^{2m-2}k^{2m} v_k^2 +{q''\over q}v_k^2\right ]. 
\ee
The equation of motion derived from the above action in Fourier space is
\be \label{EOMhighgrad}
v_k''+(\bar{\lambda}^{2m-2}k^{2m}-{q''\over q})v_k=0. 
\ee
The argument provided in previous section for the rate of the growth of $q$ and $|q''/q|$ in time, can be applied here as well. Therefore, the simplest analytical functions compatible with these conditions are  $q\sim \alpha (-y)^\mu$ with $\mu <1/2$.  This will lead to the equation of motion taking the form of a transformed Bessel equation
\be
v_k''+\left [\bar{\lambda}^{2m-2}k^{2m}-{\mu(\mu-1)\over y^2}\right ]v_k=0. 
\ee
The solutions to this equation are of the form: $C~\sqrt{|y|}~f_\nu(\bar{\lambda}^{m-1}k^my)$ where $f_\nu$ is a Bessel function of order $\nu$.  Note that Bessel equation is symmetric under $\nu\rightarrow -\nu$, so we choose the positive value of the order to represent the solutions
\ba \label{defnu}
\nu
&=&|\mu-1/2|.
\ea 
  Therefore, we can write the general solution as: 
\be \label{generalsol}
v_k(y)=C_1 \sqrt{|y|}H^{(1)}_\nu(\bar{\lambda}^{m-1}k^my)+C_2 \sqrt{|y|}H^{(2)}_\nu(\bar{\lambda}^{m-1}k^my),
\ee
where $H^{(1)}_\nu$ and $H^{(2)}_\nu$ are Hankel functions of the first kind and second kind respectively. 
 At early times, when $|\bar{\lambda}^{m-1}k^my|\gg \sqrt{ \mu(\mu-1)} = |\nu^2-1/4|^{1/2}$ and the dispersion relation is given by  $\omega_k\sim \bar{\lambda}^{m-1}k^m$, the solution should asymptote to the adiabatic vacuum condition from Eq. (\ref{vacuum}), 
\be
v_k(y)\sim \sqrt{1\over \bar{\lambda}^{m-1}k^m}\exp{[-i\bar{\lambda}^{m-1}k^my]}.
\ee

Now looking at the asymptotic behaviour of Hankel functions at early time, $|\bar{\lambda}^{m-1}k^my|\gg |\nu^2-1/4|^{1/2}$ we get
\ba
H^{(1)}_\nu(\bar{\lambda}^{m-1}k^my)&\sim& \sqrt{-2\over \pi \bar{\lambda}^{m-1}k^my}~\exp{[-i \bar{\lambda}^{m-1}k^my]}\\
H^{(2)}_\nu(\bar{\lambda}^{m-1}k^my)&\sim &\sqrt{-2\over \pi \bar{\lambda}^{m-1} k^my}~\exp{[i \bar{\lambda}^{m-1}k^my]},
\ea
where we  used the fact that $y<0$ and discarded the phase. Once again, the implication of this for our general solution (\ref{generalsol}) is that $C_2=0$ and $C_1=\sqrt{\pi\over2}$. 

We are now ready to calculate the power spectrum using the asymptotic behaviour of $H^{(1)}_\nu(l_p^{m-1}k^my)$ in the Infrared limit where $|\bar{\lambda}^{m-1}k^my|\ll |\nu^2-1/4|^{1/2}$, 
\ba 
{\cal P}^\zeta_k&\equiv&
k^3{ |v_k^2|\over M_{pl}^2 q^2} \nonumber \\
&=&k^3 \left[ {-\pi \over 2} y (H^{(1)}_\nu(\bar{\lambda}^{m-1}k^my))^2\right ]({ 1\over M_{pl}~\alpha~ (-y)^\mu})^{2}\nonumber \\
&\simeq& k^3 \left[ {\pi \over 2M_{pl}^2\alpha^2} \right ](-y)^{1-2\mu}\left[{\Gamma(\nu)\over \pi}\left({-2\over \bar{\lambda}^{m-1}k^m y}\right)^\nu\right]^2 ~~~{\rm for} ~~~|\bar{\lambda}^{m-1}k^my|\ll 1. \nonumber\\ \label{PSpec}
\ea
The spectral index for this spectrum is 
\be 
n_s-1=3-2\nu m,
\ee
so to have a near scale invariant spectrum, $\nu$ has to satisfy 
\be
\nu\simeq {3\over 2m}. 
\ee
As we had expected the standard case of $m=1$ and $\nu=3/2$ is consistent with this result.  This result also shows that since $\nu>0$, a negative value of $m$ cannot lead to a scale invariant power spectrum\footnote{A scenario with negative value of $m$ is a strange case, where smaller wavelengths cross the freezing horizon earlier than larger wavelengths.}. Also notice that, have we started with an operator corresponding to $f(k)$, instead of $k^{2m}$, then we would get 
\ba 
{\cal P}^\zeta_k \propto k^3 f(k)^{-\nu}, \ea
which could only lead to scale invariance if $f(k)\propto k^{3/\nu}$ or as we promised a power of $k$.

Again, Eq. (\ref{defnu}) will result in two solutions for $\mu$ 
\ba \label{conditionmu}
\mu_1&=&{1\over 2}+\nu={1\over 2}+{3\over 2m}\\
\mu_2&=&{1\over 2}-\nu={1\over 2}-{3\over 2m}.
\ea
However, as it was pointed out only a value of $\mu< 1/2$ corresponds to an asymptotically  constant amplitude for $\zeta$. We can see that for $\mu_1$, by substituting it in Eq. (\ref{PSpec}) and that 
\be
{\cal P}^\zeta_k\propto |y|^{-4\nu}
\ee
diverges as $y\rightarrow 0$.  In the case of $\mu_2$, ${\cal P}^\zeta_k$ is not time dependent and agrees with conservation of $\zeta$ on large scales.

The situation for $m=3$ is very peculiar. In this case $q$ is constant and the mass term in Eq. (\ref{EOMhighgrad}) drops out completely, leading to pure oscillatory solutions. This is similar to behaviour of perturbation in radiation dominated background.  In both cases, strictly speaking $\zeta$ is not asymptotically constant in Infrared limit. However, in that limit, in this case $\bar{\lambda}^{2}k^3 \ll |1/y|$, oscillations become very slow compared to  time scales of the evolution of the background. Thus, $m=3$ and $\mu=0$  describe scenarios where setting adiabatic vacuum initial condition corresponds to scale invariant power spectrum on all scales.  An example of a similar situation has been previously studied  in \cite{Mukohyama:2009gg}, motivated by the power-counting renormalizibility of quantum gravity at a Lifshitz point \cite{Horava:2009uw}.    

So far, we can conclude any succesful scenario must correspond  to 
\be
\mu\approx {1\over 2}-{3\over 2m}, 
\ee
with $m>0$.

We will now repeat the steps from inequality (\ref{ineq0}) to  inequality (\ref{ineq3}) for general case of $m >0 $.  Here instead of the lower bound of $1000$ for the ratio of largest wavelength to smallest one, we will use $e^N$ so the analogy of our final constraints to the inflationary e-folding number, $N_e$ is more clear. 
Note that, in general, freezing occurs at $\bar{\lambda}^{m-1}k^m|y|\sim |\nu^2-1/4|^{1/2} $ or $|y|\sim  {1\over2} \bar{\lambda}^{1-m}\sqrt{|9/m^2-1|} \lambda^m$. So our inequalities $\lambda_i > e^{N} ~\lambda_f$ and $\lambda_f\sim R_H(z_*)$ become
 \ba 
 |y_{i}|&\gtrsim& e^{mN} ~|y_f |, \\
 |y_f |&\sim &  {1\over2} \bar{\lambda}^{1-m}\sqrt{|9/m^2-1|} ~ R_H^{m}(z_*). 
 \ea
Now, combining these estimates, we arrive at:
  
 \be 
\label{YR}
y_f-y_i> {1\over2} \bar{\lambda}^{1-m} \times 10^{4m} (e^{mN}-1)\sqrt{|{9\over m^2}-1|} ~ R_H^{m}(z_{BBN}) .
 \ee
After Integrating the continuity equation $\dot{\rho}/\rho=-2\epsilon H$ over time and in the absence of accelerated expansion, the inequality (\ref{densitytime}) remains the same. 
If we define the average of ${b(\tau) }$ through 
\ba
 {\bar b}&\equiv&{ \int_{\tau_i}^{\tau_f} b(\tau) d\tau \over \tau_f-\tau_i} \nonumber \\&=&{y_f-y_i\over\tau_f-\tau_i},  \label{bbar}
\ea
then we have 
\be \label{maininequality2}
\ln{{\rho_i\over\rho_f}} > 2 R^{-1}_H(z_{BBN})\epsilon_{\rm min} \frac{y_f-y_i}{\bar b}.
\ee
Combining inequality  (\ref{YR}), (\ref{maininequality2}) and $\epsilon\geq 1$ leads to 
 \be\label{inequal2b}
 \ln{{\rho_i\over\rho_f}} > 10^{4m} (e^{mN}-1)\sqrt{\left|{9\over m^2}-1\right|} \frac{\bar{\lambda}^{1-m}R_H^{m-1}(z_{BBN})}{\bar b}.
\ee
Note that the dispersion relation for  angular frequency of the solution in the limit of $\bar{\lambda}^{m-1}k^m|y|\gg |\nu^2-1/4|^{1/2} $, is not generally linear in wave number. Therefore,  ${\bar b}$ does not necessarily represent the speed of sound either. However, we can use the group velocity, $c_g$,  to estimate the speed at which wave packets propagate.  Group velocity can be derived from the dispersion relation, $\bar{\omega}_k$, for $\tau$, 
\ba
c_g(\tau, \lambda)&=& {\partial \bar{\omega}_k(\tau) \over \partial k} \nonumber\\
&=& {\partial \over \partial k} \left ( {\partial \over \partial \tau}  \bar{\lambda}^{m-1}k^{m}y(\tau)\right )\nonumber\\
&=&b(\tau) m \bar{\lambda}^{m-1}k^{m-1}\nonumber\\
&=&b(\tau) m \left ({\lambda\over \bar{\lambda}}\right )^{1-m}.
\ea

\begin{figure}[htpb]
\includegraphics[width=\columnwidth]{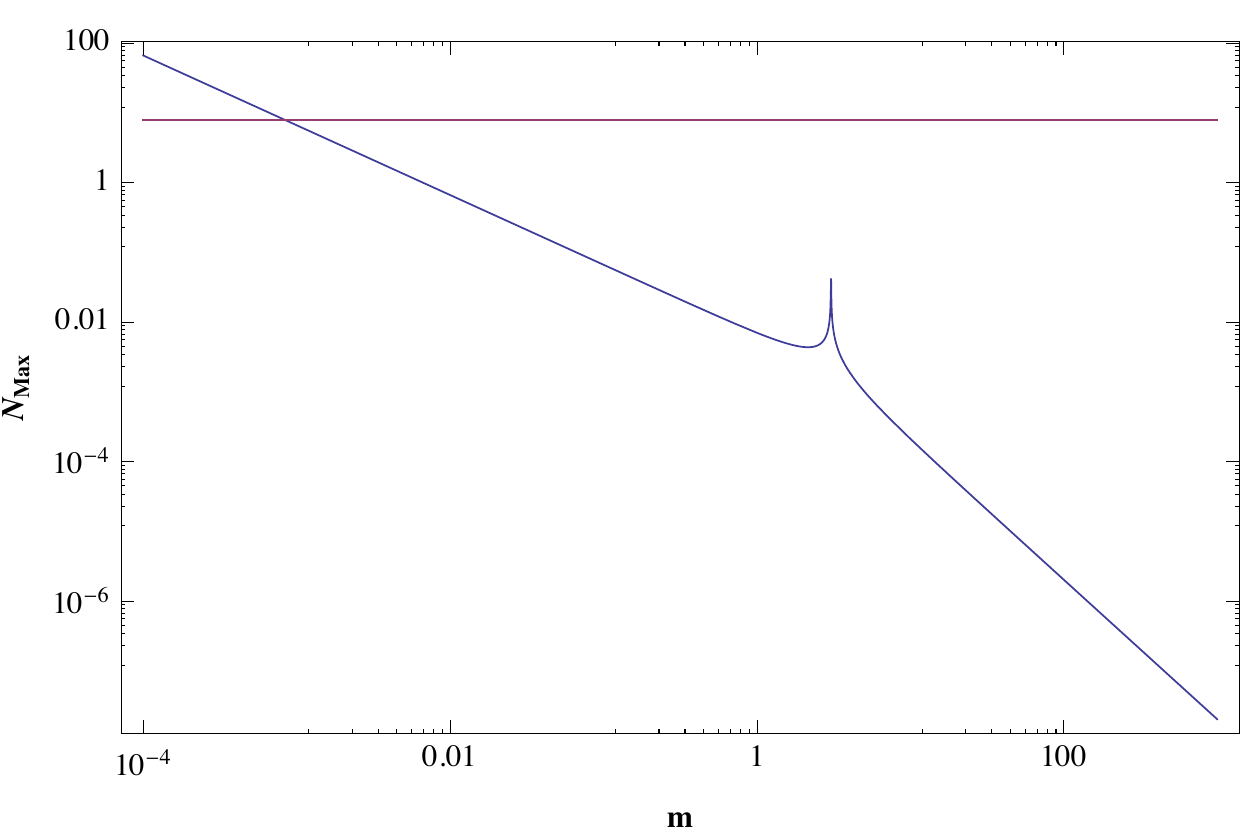}
\caption{Upper bound on the number of scale-invariant, sub-luminal, and sub-planckian e-foldings, $N_{\rm max}$,  for different values of $m$, the ``number'' of spatial derivatives of curvature in the quadratic action. $N_{\rm max}(m)$ diverges for $m \rightarrow 0$ and $3$. It should be $\gtrsim 8$ to be consistent with cosmological observations. }\label{Nbound}.
\end{figure}

If we define $c_{\rm max}(\lambda)$ as the maximum value of $c_g(\tau, \lambda)$ during the period of $\tau_i$ to $\tau_f$,  we obtain 
\be 
{\bar b}<m^{-1}\left ({\lambda\over \bar{\lambda}}\right )^{m-1}c_{\rm max}(\lambda).
\ee
Substituting this result in relation (\ref{inequal2b}) leads to 
\be
c_{\rm max}(\lambda) \ln{{\rho_i\over\rho_f}} >10^{4m}  (e^{mN}-1)\sqrt{|9-m^2|}  \left(\frac{R_H(z_{BBN})}{ \lambda} \right)^{m-1} . \label{inequal3b}
\ee
While this constraint could be applied for any wavelength $\lambda_f< \lambda <\lambda_i$, the notion of signal propagation is only relevant for scales smaller than the freezeout horizon (i.e. in the WKB regime), where group velocity characterizes the speed of signal propagation.  Since only the shortest wavelength $\lambda_f$ remains in this regime throughout the period $ \tau_i < \tau < \tau_f$, Eq. (\ref{inequal3b}) can only be used to limit the speed of signal propagation for $\lambda = \lambda_f$.  Substituting $\lambda_f \sim1~ {\rm Mpc} \sim 10^{4} R_H(z_{BBN})$ \footnote{Note that $\lambda$ is in comoving scales so its value is the same as today.} in the inequality (\ref{inequal3b}) will require
\ba\label{inequal4}
c_{\rm max} \ln{{\rho_i\over\rho_f}} > 10^{4} (e^{mN}-1)\sqrt{|9-m^2|}. 
\ea
This result demonstrates how the effects of  sub/superluminality, the allowed range of energy density and number of e-foldings have to compensate each other in order to satisfy the observational constraints. For example, for both cases of  $m=1$ and $m=2$ substituting the observational evidence of $N>8$ ($\lambda_i/\lambda_f \gtrsim 3\times 10^3$) implies either superluminality or super-Planckian energies.  In fact, requiring only the range allowed by subluminality and sub-Plankian energy densities  ($c_{\rm max} \ln{{\rho_i\over\rho_f}} \lesssim \ln{{M^4_{pl}\over\rho_{BBN}}} \sim 88\ln 10$),  we can obtain an upper bound on the value of $N$ 

\ba \label{Nresult}
N< {1\over m}\ln \left[ {2.03\times 10^{-2}\over \sqrt{|9-m^2|}}+1\right].
\ea
Figure (\ref{Nbound}) shows the trend of upper bound on $N$ for different values of $m$. 
This bound rules out $N\sim 8$ for any (positive) $m$, with the exception of $|m-3| \lesssim 10^{-25}$, or $m \lesssim 10^{-3}$. As we expected, the analysis breaks down at $m=3$ (where we have Lifshitz symmetry).

To summarize this section, most local and non-local actions with nontrivial spatial terms cannot admit alternative scenarios to inflation without violating some principles of general relativity or are very tightly constrained by observations. We first showed that the Fourier transform of these terms should have a power law form with a positive power to admit scale invariant spectrum. Next, we showed that no value of $m$, except for $m\simeq 3$ or $0<m\lesssim 10^{-3}$ (the latter corresponds to very non-local operators),  supports the observed data in the framework of general relativity, without an accelerated expansion phase (or inflation).

Note that the argument expressed above relies on the validity of the freezing process to turn vacuum perturbations into a scale-invariant spectrum. However, for $m\approx 3$ since Vacuum perturbations are scale-invariant to begin with, this argument can not be applied. Also in this case choosing a different value of $\mu$ that does not comply with Eq.(\ref{conditionmu}), does not necessarily rule out the scenario. A none zero $\mu$ may only set an upper bound on scales bellow which spectrum is scale invariant. So as long as that scale is larger than $(a_0H_0)^{-1}$ the spectrum is consistent with the constraints. 

It is also worth mentioning that including a tilt, i.e. the fact that the power spectrum does not appear to be exactly scale-invariant can easily be implemented in this analysis by inserting $n_s-1\sim -0.04$. That will slightly change the relation between $N_{\rm max}$ and $m$\footnote{$\sqrt{|9-m^2|}$ in Eq. (\ref{Nresult}) has to be replaced by $\sqrt{|3.04^2-m^2|}$.}. However, it will not make any significant impact on our conclusion.  For example, the value of $m$ for which $N_{max}$ diverges will be shifted to $m \simeq 3.04$ but the plot in Figure (\ref{Nbound}) remains essentially the same. 

 Furthermore, the derivation of the upper bound on $N$ for $k^m$ operators is nearly independent of the choice of initial condition. Basically, the vital information about behaviour of the solution that was used to derive the bound came from the dispersion relation in the ultraviolet regime, and that the modes have to cross a freezing scale.  However, this information can be derived from equation of motion and does not depend on initial conditions. We showed that choosing adiabatic vacuum initial condition fixes the relation between $\nu$ and $m$ through Eq. (\ref{PSpec}). Choosing a different initial condition introduces, $C(k)$, some function of $k$ in RHS of Eq. (\ref{PSpec}) which needs to be canceled off by $k^{2m\nu-3}$. This is only possible if  $C(k)$, is a power of  $k$, i.e. $C(k) \propto k^s$. Substituting $2m\nu-3=s$ into our analysis will consequently lead to replacing  $\sqrt{|9-m^2|}$ in Eq. (\ref{Nresult}) by $\sqrt{|(3+s)^2-m^2|}$ which only stretches or squeezes Plot (\ref{Nbound}) depending on the power, $s$, but does not affect the behaviour of $N_{\rm max}$ qualitatively.  This enables us to extend our result to different choices of initial conditions that could also lead to scale-invariance. 

\section{Actions with mixed temporal and spatial derivatives} \label{time spatial} 

In the previous section, we studied a large class of actions, which involved non-trivial spatial derivative operators. However, one can ask wether the same analysis can be extended to operators which have mixture of time and spatial derivatives. In general, the inclusion of such terms  can change the nature of equation of motion drastically. For example, we can no longer benefit from a Bessel type equation of motion. This means complexities in regard to possibility of setting adiabatic vacuum conditions, distinguishing a freezing scale and eventually obtaining a scale invariant spectrum.  In this section, we consider a simple example of such actions to demonstrate some of these subtleties and why, in general, we do not expect such actions to lead to a scale-invariant spectrum.  The following action involves a term with both time and spatial derivative

\be \label{action_mixed}
S_2={M_{pl}^2\over 2}\int d^3x~ d\tau z^2 \left (\tau \right ) \left [  \left ( \partial\zeta \over \partial\tau \right )^2 +  \bar{\lambda}^{2m}b^2 \left (\tau \right )  \left ( \nabla^m  {\partial\zeta \over \partial \tau} \right )^2 \right ].
\ee
As before, $b \left (\tau \right )$ is a dimensionless function of $\tau$. Expressing the above action in Fourier space we obtain
\ba \label{action_zeta_k}
S_2 &=& {M_{pl}^2\over 2} \int d^3k d\tau z^2 \left (\tau \right ) \left (  1+k^{2m} {b^2}(\tau) \bar{\lambda}^{2m}  \right) \zeta_k^{\prime 2} \left (\tau \right )\nonumber \\
    &=& { {M_{pl}^2} \over 2} \int d^3k d\tau q^2 \left (k,\tau \right)\zeta_k^{\prime 2}(\tau),
\ea
where prime in this section represents $\frac{\partial}{\partial \tau}$ and $q$ is given by 
\be
q^2 \left (k,\tau \right) \equiv z^2 \left (\tau \right ) \left[  1+\bar{\lambda}^{2m} k^{2m} {b^2}(\tau) \right].
\ee
Defining the canonically normalized variable $v_k \equiv M_{pl} ~q ~\zeta_k$, the action given in \ref{action_zeta_k}, expressed in terms of the new variable $v_k \left (k,\tau \right )$, takes the following form
\be
S_2= {1 \over 2} \int d^3k d\tau \left [ v_k^{\prime 2} + {q^{\prime\prime} \over q} {v_k}^2  \right ],
\ee
and the corresponding equation of motion would be
\be  \label{equation of motion nu}
v_k^{\prime\prime}- {q^{\prime\prime} \over q} {v_k} = 0.
\ee
This equation has general solutions  given by
\be \label{solutionv}
v_k = C_1 (k ) q (k,\tau) + C_2 (k) q(k,\tau) \int \frac{1} {q^2(k,\tilde{\tau})} d\tilde{\tau}.
\ee
Next, we need to examine whether there exists any particular solution which is consistent with an attractor FRW background, imposing adiabatic vacuum condition at early times, and leads to scale-invariance at late times. This means the time dependence of $\int \frac{1} {q^2(k,\tau)} d\tilde{\tau}$ is decaying in time, there is a limit where $w\sim \sqrt{ {q^{\prime\prime} \over q}}$ satisfies WKB conditions and finally, $C_1(k )^2\propto k^{-3}$. 
We assume different hypothetical behaviours for $b(\tau)$ and argue that all these three conditions can not be satisfied at the same time. 

The first case is when in early times $1\ll \bar{\lambda}^{2m} k^{2m} {b^2}(\tau)$ for the wavelengths of our interest, such that  $q\sim z~b~ \bar{\lambda}^{m} k^m $.  In this case, if the WKB regime exists it must be described by $w\sim \sqrt{ -{(zb)^{\prime\prime} \over z b}}$ which is independent of $k$. Since the adiabatic vacuum solution should match the solution given by Eq. (\ref{solutionv}) in the same limit, we conclude $C_1(k )\propto k^{-m}$ and $C_2 (k )\propto k^{m}$, in order to cancel the $k$ dependences of each term. However,  $C_1(k )\propto k^{-m}$ together with 
 $C_1(k )^2\propto k^{-3}$ already rule out the possibility of scale-invariance for any $m\neq 3/2$.  
  
Next we assume in early times $1\gg \bar{\lambda}^{2m} k^{2m} {b^2}(\tau)$. Similarly in this case, if the WKB regime exists it must be described by $w\sim \sqrt{ {z^{\prime\prime} \over z }}$ which is independent of $k$ as well. Comparing the vacuum solution with Eq. (\ref{solutionv}) we conclude $C_1(k )$ and $C_2 (k )$ are also independent of $k$. However, that is inconsistent with the condition $C_1(k )^2\propto k^{-3}$. 
One could imagine other possibilities for the asymptotic behaviour of $q$ in the early time as well. However, in general  satisfying a WKB initial condition where $v_k\sim \sqrt{q\over q''} \exp{[{- i \int \sqrt{q''\over q} d\tau}]}$ and at the same time Eq. (\ref{solutionv})  in that limit with $C_1(k )^2\propto k^{-3}$ is not possible. 

To conclude this section, inclusion of terms with mixture of time and spatial derivatives changes the nature of equation of motion drastically, which makes them harder to investigate without knowing the details of the scenario. However, it seems very unlikely that these types of actions could be successful as alternatives for generating scale-invariant perturbations. For example, for actions of type Eq. (\ref{action_mixed}) and with a very limited knowledge about the time dependence of $b(\tau)$, we could rule out the possibility of generating scale-invariant power spectrum, except for when $m=3/2$. That is not to say that  the non-local case of $m=3/2$ will naturally lead to scale-invariance either. Even in that case, engineering the time dependence of both $z(\tau)$ and $b(\tau)$ to satisfy all the constrains is not a feasible task\footnote{We postpone the study of these unusual scenarios to another paper.}.
\section{Concluding Remarks} \label{conclusion} 
Cosmological observation of Cosmic Microwave Background (CMB), as well as Large Scale Structure (LSS) both point to  initial conditions consistent with adiabatic, nearly scale-invariant and Guassian primordial spectrum of perturbations. The state of present observations can attest that these features persist for at least three decades of wavelengths in the sky. Combining this simple piece of information with our other observationally tested knowledge about early universe, i.e. that the scale of observed perturbations were about four orders of magnitude larger than Hubble radius during Big Bang nucleosynthesis, enables us to significantly constrain the possible scenarios of early universe. 

In this paper we studied this question, assuming such perturbations were generated by including higher order derivatives or non-local operators in the quadratic action for curvature perturbations $\zeta$. This can be considered as a natural generalization 
of a standard action with only first order derivative for $\zeta$. Under the assumption that these 
fluctuations were generated from adiabatic vacuum initial conditions, we were able to deduce important constraints on the background evolution and validity of the different tenets of general relativity. 
 
 Our result was an improvement and extension of the domain of validity of the theorem proved in \cite{Geshnizjani:2011dk}, to a far larger class of  local and nonlocal actions. This theorem states that any scenario which does not have a phase of accelerated expansion (or inflation) must satisfy one or more of the following conditions: violating Null Energy Conditions (NEC), superluminal speed of sound, or super-Planckian energy densities. We showed that without breaking any of these conditions, a large group of alternative scenarios for inflation with higher derivative or non-local actions are ruled out. Interestingly, only theories in the neighbourhood of Lifshitz points with $\omega_k \propto k^0$ and $k^3$ remain viable. 

It is also worth mentioning that our result is nearly independent of the choice of initial conditions. Basically, the vital  information about behaviour of the solution were the dispersion relation in the ultraviolet regime and the fact that modes  crossed a freezing scale. However, this information can be derived from the equation of motion and does not depend on initial conditions/state. An initial condition different from adiabatic vacuum can change the value of Lifshitz point $\omega_k \propto k^3$, where the exclusion neighbourhood is located, to a different point but it does not affect our result qualitatively. This enables us to extend our result to different choices of initial conditions that could also lead to scale-invariance. 

\section*{Acknowledgements}
The authors would like to thank Will Kinney and Azadeh Moradinezhad Dizgah for
invaluable discussions which inspired this work. NA is grateful to University
of Tehran for supporting this project under a research grant provided by the university research council. She also thanks Perimeter Institute for Theoretical Physics for their hospitality during her visit in 2012-2013 where much of this work was performed. GG research is supported by the Discovery Grant from Natural Science and Engineering Research Council of Canada, the University of Waterloo and the Perimeter Institute for Theoretical Physics. Research at the Perimeter Institute is supported by the Government of Canada through Industry Canada and by the Province of Ontario through the Ministry of Research $\&$ Innovation. 
\bibliographystyle{JHEP}
\bibliography{SCINC.bib}
\appendix
\section{Producing Scale Invariant spectrum in any expansion history} \label{generalbg}
In Section \ref{setup}, we demonstrated that a cosmological background satisfying $q\propto y^{-1}$ can lead to scale invariant superhoriozn curvature fluctuations. For scenarios sourced  by scalar fields or hydrodynamical fluids, $q$ is related to the expansion of the universe and the speed of sound through:
\be
q=\frac{a \sqrt{2\epsilon}}{\sqrt{c_s}}.
\ee
Combining these two conditions, we get 
\be
\label{beta0}
\frac{a \sqrt{2\epsilon}}{\sqrt{c_s}}=\frac{-\alpha}{y}, 
\ee
where $\alpha$ is a positive number and related to the amplitude of perturbations, through Eq. (\ref{PS1}).
The equation above, as we show here, can be easily solved for any expanding background. 
We start be rewriting $y$ in terms of conformal time, $\tau$ in Eq. (\ref{beta0}), 
\be
\label{aepscsalphay}
{a\sqrt{2 \epsilon}\over \sqrt{c_s}} = {\alpha \over \int_\tau^{\tau_e}  c_s d\tau},
\ee
where $\tau_e$ refers to the corresponding conformal time at $y_{end}=0$. This equation leads to a first order differential equation for $c_s$:
\be
{-c_s\over \alpha} ={(\sqrt{c_s})_{,\tau}-\sqrt{c_s}(\ln{a}\sqrt{2\epsilon})_{,\tau}\over a\sqrt{2 \epsilon}}.
\ee
By a change of variable $u\equiv 1/\sqrt{c_s}$, we get 
\be
{a\sqrt{2 \epsilon}\over \alpha }\left ({1\over u}\right )- {u_{,\tau}\over u} =(\ln{a}\sqrt{2\epsilon })_{,\tau}, 
\ee
which can be re-organized as 
\be \label{ss}
(ua\sqrt{2 \epsilon})_{,\tau}= {2 a^2\epsilon\over \alpha}. 
\ee
Now integrating (\ref{ss}) we obtain:
\be \label{csfrombackground}
 u(\tau)={\sqrt{2}\over \alpha a(\tau) \sqrt{\epsilon(\tau)}}\int a^2(\tilde{\tau}) \epsilon(\tilde{\tau}) d\tilde{\tau}.
\ee
Therefor for any given expansion evolution described by $a(\tau)$, one can derive the speed of sound, $c_s(\tau)$ such that the scale-invariant condition is satisfied. The followings are some known examples which can be derived from above relation:
\begin{itemize}
 \item Taking inflationary or adiabatic ekpyrotic cases that $a(\tau) \sqrt{\epsilon(\tau)} \sim -1/\tau$ (note that $\tilde{\tau} \in [-\infty , \tau])$ leads to  $c_s\sim const.$.
 \item Taking $\epsilon \sim const.$ we have $a\propto \tau^\gamma$ ($\gamma\equiv 1/(\epsilon-1)$). Therefore from Eq. (\ref{csfrombackground}), $u\propto \tau^{\epsilon/(\epsilon-1)}$ and leads to $c_s\propto \tau^{2\epsilon/1-\epsilon}$ . In particular setting  $a\propto \tau$ or $\epsilon=2$ ($\tau \in [0, +\infty]$) results in $u\propto \tau^2$ hence $c_s\propto \tau^{-4}$. This case is referred to as tachyacoustic model where sound speed starts very large and falls rapidly (see \cite{Bessada:2009ns} for more details). Also note that in this case $y\sim -1/\tau^3$ and even though $\tau$ runs from zero to infinity, $y$ runs from minus infinity to zero consistent with our horizon exiting argument put forward before. We also get a similar result for $a\sqrt{\epsilon}\propto \tau^\gamma$, including cases that $\epsilon$ is varying rapidly but variation of $a$ or $H$ are not significant. 
\end{itemize}

Once $a(\tau)$ and $c_s(\tau)$ are known, one can always reconstruct non-canonical scalar field actions consistent with their evolution \cite{Bean:2008ga}.  


\section{Disambiguation of different cosmological scales in scenarios with varying speed of sound} \label{apxHorizonz}
The significance of different Cosmological horizons/scales is often confused in the literature. This is mostly because some of the common assumptions that govern the dynamics of the cosmological models, dictate particular relations between different scales as well. The confusion has been further enhanced by the fact that, historically, these coincidental relations have lead the cosmologists to use misleading terminology as well. However, it is important to keep in mind that once dealing with models that deviate from the standard cosmological scenarios, as some of the underlying assumptions are removed, cosmological  scales may not inherit the characteristics we are familiar with either.  A famous example of this phenomenon is of course the well known resolution to the {\it Horizon} problem in inflationary scenarios.  Unlike radiation dominated universe, during inflation the Hubble radius and causal horizon deviate significantly from each other, making it possible for causal horizon to grow exponentially (in physical coordinates), while the Hubble radius remains nearly  constant.  

The situation becomes more subtle when sound speed deviates from unity, and furthermore,  varies in time significantly. In such cases, the relevant scales for influence of background gravitational effects, or the freezing process, will no longer be given by the Hubble radius. In what follows, we briefly review the definition of these scales to clear up some of the misconceptions about these scales. 

We start by reminding the readers about the significance of the {\it Causal horizon}. This scale often arises in connection to the topic of horizon problem. The thermal homogeneity of CMB across the sky points to a common initial condition on very large scales. There is an issue, however, if we assume a causal mechanism starts homogenizing our observable universe at some finite time in the past, $\tau_{in}$. For an FRW metric, the size of this scale is simply given by the distance light has traveled since that initial time, which in comoving coordinates corresponds to:
\ba
R_L&\equiv& \int_{\tau_{in}}^\tau d\tau = \tau- \tau_{in}. 
\ea
For scenarios with varying speed of sound, we can also calculate this radius in terms of the time variable $y$ through the following relation
\ba
R_L\equiv \int_{y_{in}}^y {dy\over c_s(y)}.
\ea
In the old standard cosmology,  universe was dominated either with matter or radiation. In this picture, taking the initial time to be when the energy density is Planckian or even when it diverges i.e. the {\it Singularity}, would result in a small casual horizon not sufficient to cover our observable Universe, which leads to the {\it Horizon} problem.  However, during inflation, since energy density is almost constant $\tau_{in}$ can be pushed significantly further back\footnote{This is a very simplified version of the story. There are some subtle issues regarding initiating the inflation, Bunch-Davies vacuum initial conditions, etc. These issues can complicate the problem and even question whether inflation actually resolves the Horizon problem. However, those discussions are beyond the scope of this paper.}.

Next, let us move on to the so called {\it Hubble radius} or as it is sometimes referred to {\it Hubble horizon}. This scale, as its name suggests, simply characterizes the length scale calculated from Hubble constant, and in comoving coordinates is given by: 
\ba
R_H\equiv {1\over aH}.
\ea
In the old standard cosmology the value of causal horizon and this scale follow each other closely. However, as we pointed out earlier, the two of them differ significantly for slow-roll inflationary scenarios. Causal Horizon expands in time during slow-roll inflation (both in physical and comoving coordinates) but Hubble constant remains nearly constant and $R_H$ shrinks during this period. An important relevance of this scale for single field slow-roll models is that it also represent the scale of space-time curvature; The radius of curvature for these models  translates into the size of the wavelengths where modes start to freeze out.  

 As we explained in Section \ref{setup} in detail, the scale of freezing (if the model allows it) can be directly approximated by Eq. (\ref{Rz}):
\be
R_\zeta\equiv \sqrt{q\over q''}
\ee
and it often but not necessarily also coincides with the scale that WKB approximation breaks down. However, we know $ \zeta_k$ does not become constant in the presence of entropy perturbations. Furthermore, there are cases where $R_\zeta$ diverges, such as the radiation dominated backgrounds, but there is still some finite scale above which modes are approximately conserved. Therefore, it might be more accurate to say, for adiabatic perturbations $R_\zeta$ sets an upper bound on the threshold outside which, $\zeta_k$  is almost constant. 
One can show that, for standard actions, if $c_s=1$ and slow-roll parameters are small then $R_\zeta\sim R_H$. If the assumption of  $c_s=1$ is relaxed into a constant speed of sound then $R_\zeta\sim c_s R_H$. 

Last, let us also say a few words about {\it sound horizon}, $R_C$, relevant to scenarios with non-unitary speed of sound.  Similar to causal horizon, one can wonder about the distance sound-waves have travelled since some initial time. This scale, could depend on wavelengths as well.
We can calculate this distance in the following way
\ba
R_C&\equiv& \int_{\tau_{in}}^\tau c_s(\tau) d\tau\nonumber \\&=&y-y_{in}. 
\ea
 Once again under certain assumption such as constant $c_s$ or small slow-roll variables one may find, $R_C\sim c_s R_L$,  $R_C\sim R_\zeta$ or $R_C\sim c_s R_H$, but in general these relations may not hold.  

\begin{figure}[!ht] 
\begin{center}
\includegraphics[scale=1.01]{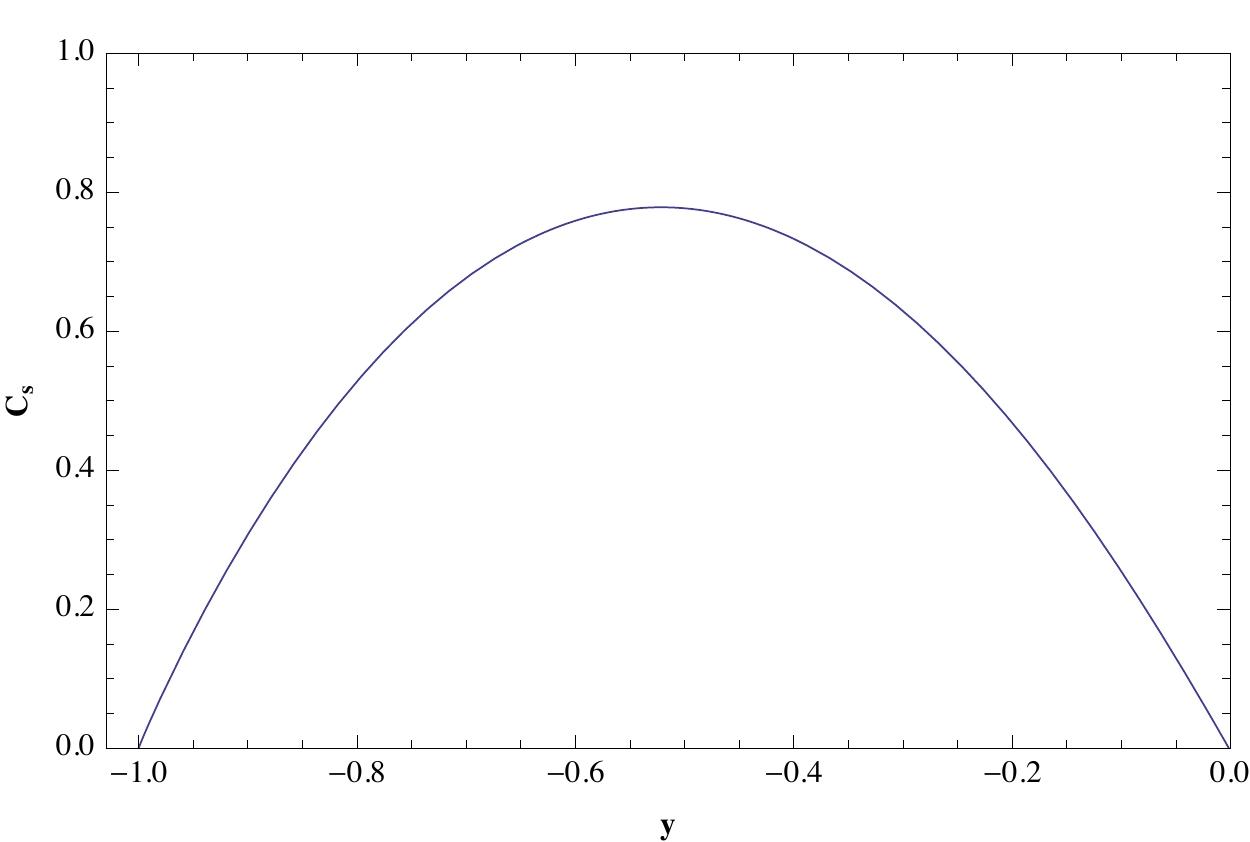}
\hspace{3mm}
 \includegraphics[scale=1.0]{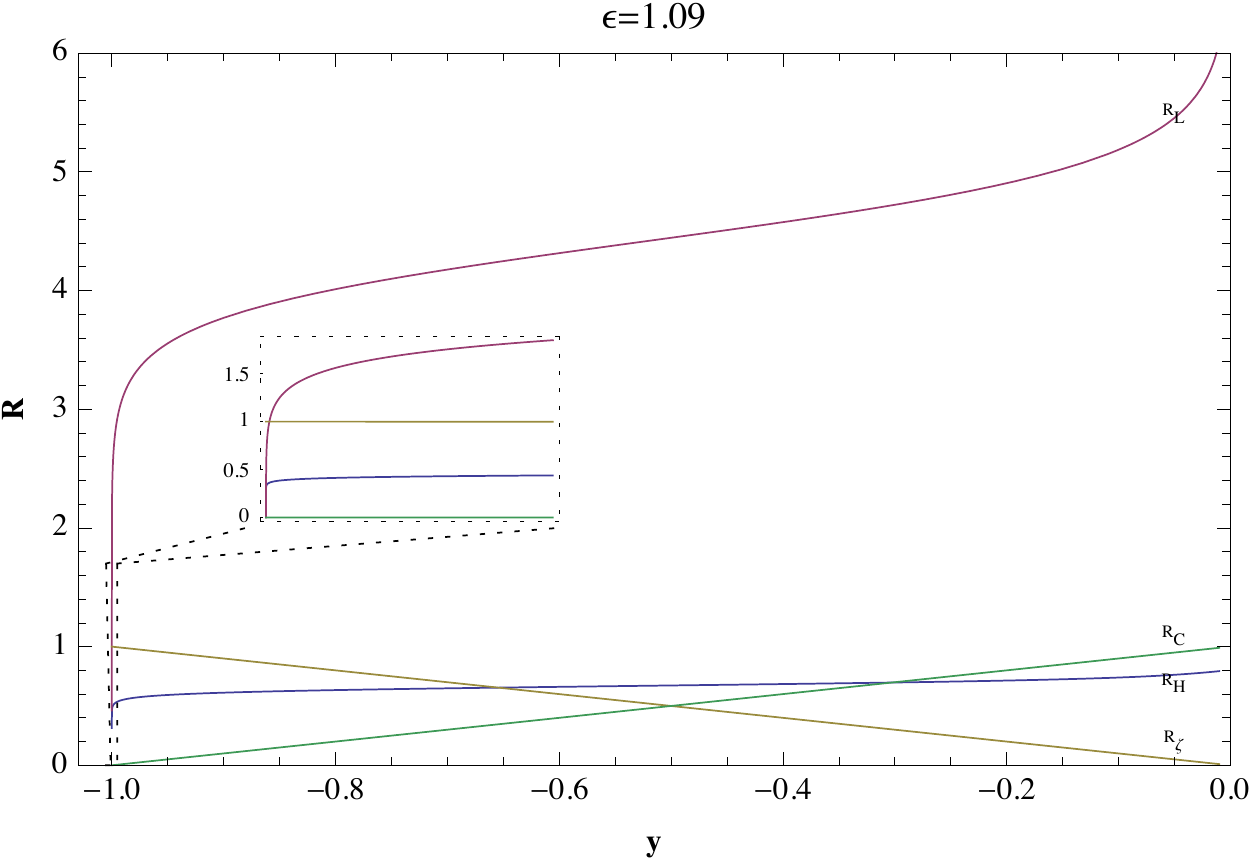}
\caption{An example of a non-accelerating universe with  $\epsilon = 1.09$,  in which sound speed varies such that  one can generate a scale invariant spectrum of curvature fluctuations. The top plot shows the evolution of sound speed and that the propagation of modes remains subluminal for this model at all times. The lower plot demonstrates the evolution of different cosmological scales as functions of $y= \int c_s d\tau$. The magenta and green curves represent the distances light and sound each travel respectively, since the Planck time. The orange and blue curves represent freezout and Hubble radii respectively. \label{horizons} }
\end{center}

\end{figure}
To end this appendix, we provide an example in which these scales behave very differently from each other.  
Using the method in Appendix (\ref{generalbg}), we were able to numerically explore the solutions in which sound speed varies but such that it can still generate scale invariant modes. For simplicity and to produce dramatic effects we chose a non-accelerating universe with constant $\epsilon=1.09$. The model represents a background which is close to zero decelerated expansion. Nevertheless, it suffices for our illustrative purposes.  We also set the initial time to when energy density is Planckian. Figure (\ref{horizons}) illustrates the result for this model. The top plot in the figure shows the derived evolution of sound speed in order to produce scale-invariant spectrum . As we see the propagation of modes remains subluminal for this model at all times. The lower plot demonstrates how the evolution of different cosmological scales differ from each other as functions of $y$.  Some interesting features to note are for example that causal horizon becomes much larger than all the other scales very early on. Another interesting feature is  that even though $c_s$ remains less than one at all times, contrary to the common belief  that $R_C< R_H$ for $c_s<1$, sound horizon crosses Hubble radius. That intuition is mostly based on the situations where $R_C\sim c_s/aH$ which is definitely not valid for cases where sound speed varies significantly.

\end{document}